\documentclass[11pt]{article}

\usepackage[a4paper,margin=2.5cm]{geometry}
\usepackage[utf8]{inputenc}
\usepackage[T1]{fontenc}
\usepackage{amsmath,amssymb}
\usepackage{graphicx}
\usepackage[hidelinks]{hyperref}
\usepackage{authblk}
\usepackage[numbers,sort&compress]{natbib}
\usepackage{microtype}

\usepackage{fancyhdr}
\title{\bfseries Estimation of the laser guide star uplink tip-tilt\\ using aperture size diversity}

\author[1,3,*]{Pierre Jouve}
\author[2]{Thierry Fusco}
\author[4,5]{Carlos M. Correia}
\author[2,1]{C\'edric Taissir Heritier}
\author[2,1]{Jean-Fran\c{c}ois Sauvage}
\author[1]{Benoit Neichel}

\affil[1]{Laboratoire d'Astrophysique de Marseille, UMR 7326, 13388 Marseille, France}
\affil[2]{DOTA, ONERA, BA 701, F-13661 Salon Air Cedex, France}
\affil[3]{Space ODT, Porto, Portugal}
\affil[4]{Faculty of Engineering, University of Porto, Porto, Portugal}
\affil[5]{INESC TEC -- Institute for Systems and Computer Engineering, Technology and Science, Porto, Portugal}
\affil[*]{Corresponding author: pierre.jouve@lam.fr}

\date{\today}

\begin{document}

\maketitle

\begin{abstract}
\noindent
Laser guide star (LGS) adaptive optics cannot directly measure tip-tilt (TT),
forcing reliance on natural guide stars and limiting sky coverage. We propose
estimating the uplink TT from telemetry acquired at the laser launch telescope
(LLT) alone, operated in a monostatic configuration as both emitter and
receiver. Extracting TT over concentric disks of different diameters within
the receiving pupil yields signals mixing uplink and downlink contributions
in different proportions; this aperture size diversity, combined with a
linear minimum mean square error (LMMSE) estimator, disentangles the uplink
component. Simulations show a residual error of 20--35 mas under median
seeing (0.65$''$) conditions.
\end{abstract}

\vspace{1em}

\section{Introduction}

The laser guide star (LGS) tip-tilt (TT) indetermination problem described in \cite{rigaut1992laser} is a long-standing issue arising from the fact that, as the laser beam propagates upward through the atmosphere, it experiences dynamic tilts induced by atmospheric turbulence. These uplink tilts are then mixed with the turbulent phase distortions affecting the downlink path, leading to ambiguity in determining the TT measured using the LGS. Beyond astronomy, this limitation is also becoming critical for ground-to-space free-space optical communications, where adaptive optics pre-compensation of the uplink beam faces the same tilt indetermination \cite{lognone2024tip,barrett2026tip}. Several approaches have been proposed to mitigate this effect \cite{ragazzoni1997robust}, with more recent developments exploring tomographic reconstruction techniques \cite{reeves2016tomographic}. The idea of using an emitter and receptor of different size was mentioned in \cite{belen1994fundamental}; we aim to further develop this concept through modern analytical tools and comprehensive simulations.

The goal of this letter is to present a new method based on aperture size diversity to estimate the uplink contribution of the LGS TT directly using the laser launch telescope (LLT) as an emitter and receptor in a monostatic configuration. We focus on the methodology and explore the potential and performance of this approach under controlled, simplified conditions, and believe the concept could reduce dependence on natural guide stars and increase the flexibility of AO observations.

\section{Principle of the method}\label{sec:method}

\subsection{Monostatic setup with aperture size diversity}

\begin{figure}[htb]
\centering
\includegraphics[width=0.65\linewidth]{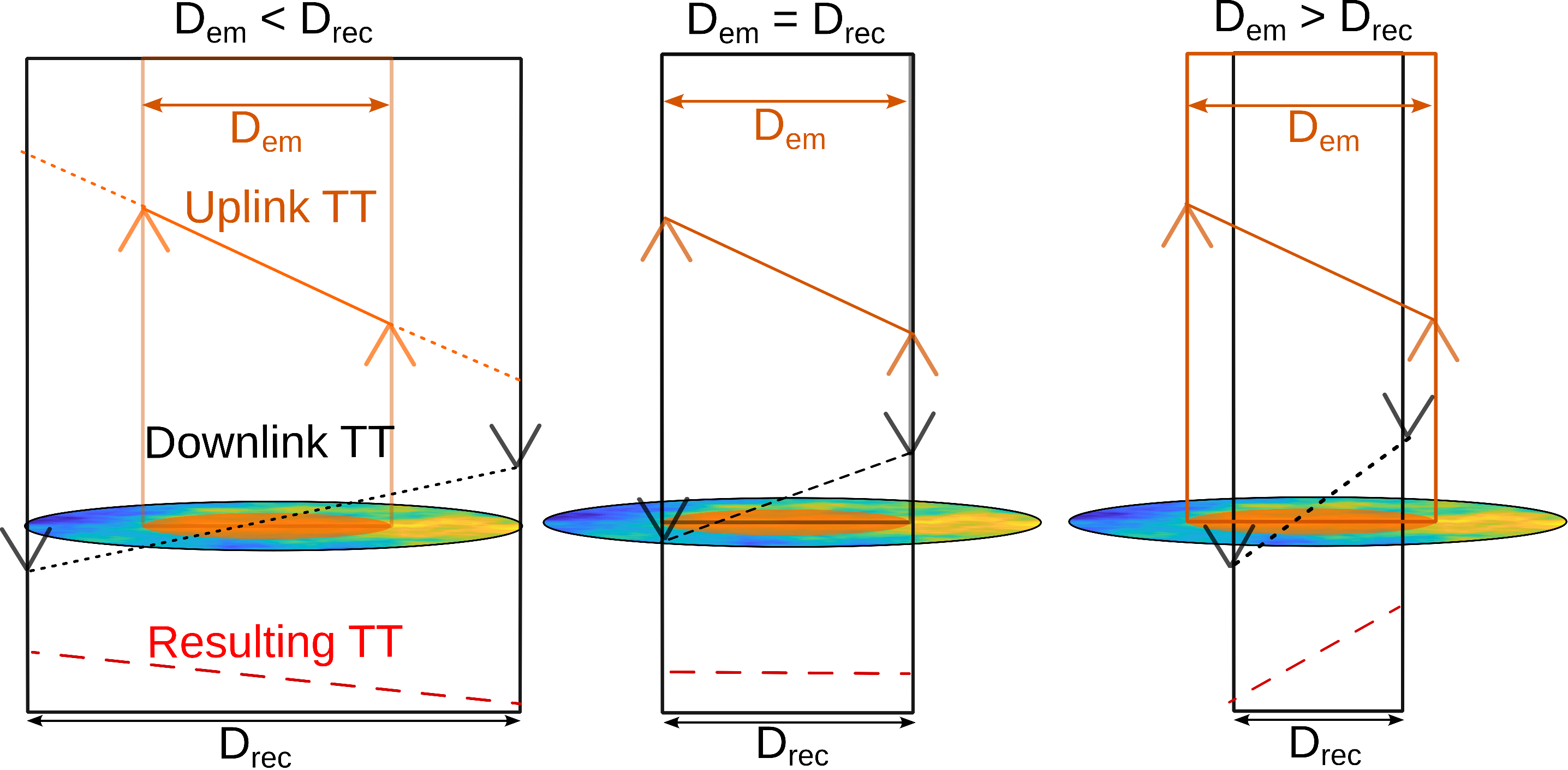}
\caption{Simplified scheme of an uplink-downlink propagation in a monostatic configuration. The uplink TT is generated by a beam of diameter equal to emitter size, and the downlink is the TT generated by a beam of diameter equal to receptor size. The receptor sees the differential between the downlink and the uplink TT contributions.}
\label{fig:scheme}
\end{figure}

We consider a monostatic situation with a fixed emitter aperture of diameter $D_\mathrm{em}$ and a varying receptor aperture of diameter $D_\mathrm{rec}$ (Fig.~\ref{fig:scheme}). We emphasize that the receiver is a dedicated aperture located at the LLT itself, concentric with the emitted beam; it is distinct from, and much smaller than, the science telescope pupil. To better isolate the dynamics of uplink and downlink TT, we adopt a simplified framework. We first assume a single atmospheric layer located at ground level and neglect the light travel time between the telescope and the LGS at 90\,km (${\sim}0.6$\,ms $< \tau_0 \sim 3$--$5$\,ms at Mauna Kea \cite{schock2009tmt}). The anisoplanatism arising from uplink tilt is neglected ($\sigma_\mathrm{TT} \sim 200$\,mas $\ll \theta_0 \sim 2$--$3''$ at Mauna Kea \cite{schock2009tmt}). The measurements are noiseless; therefore, the photon noise propagation through the LMMSE estimator is left to future work. Finally, in Sect.~\ref{sec:results}, we add a second layer and therefore take into consideration the cone effect on the downlink path.

\subsection{Statistical properties of multi-aperture tip-tilt measurements}

\subsubsection{Aperture dependence of tip-tilt variance}

The fundamental principle underlying our method is the dependence of TT angular jitter $\sigma_{\alpha}$ on telescope aperture size. The relationship between the TT variance and aperture size is given in a general context with finite outer scale $L_0$ in \cite{conan2000modelisation}, see Fig.~\ref{fig:diff}(a), where the analytical model is compared with numerical simulations performed with the OOPAO simulator \cite{heritier2023oopao}, used throughout this letter. The outer scale reduces the amplitude of $\sigma_{\alpha}$, and amplifies the contrast between small and large aperture jitter values.

\subsubsection{Cross-correlation between apertures}

We derive the spatial correlations of the TT between two concentric pupils of diameters $D_1$ and $D_2$ using the general formula from \cite{wilson1996adaptive}:
\begin{equation}
\rho_{\mathrm{TT}}(D_1,D_2)
=
\frac{
\int_{0}^{\infty}
W_{\phi}(k)\,
\frac{J_{2}(\pi D_1 k)\,J_{2}(\pi D_2 k)}{k}
\,\mathrm{d}k
}{
\sqrt{\mathcal{I}(D_1)\,\mathcal{I}(D_2)}
},
\label{eq:Wilson}
\end{equation}
where $J_2$ is the Bessel function of the first kind of order~2, $k$ is the spatial frequency, $\mathcal{I}(D) = \int_{0}^{\infty} W_{\phi}(k)\,J_{2}^{2}(\pi D k)/k\,\mathrm{d}k$, and $W_{\phi}(k) = 0.0229\,r_0^{-5/3}(k^{2}+L_{0}^{-2})^{-11/6}$ is the von K\'arm\'an power spectrum, with $r_0$ the Fried parameter.

\begin{figure}[htb]
\centering
\includegraphics[width=0.75\linewidth]{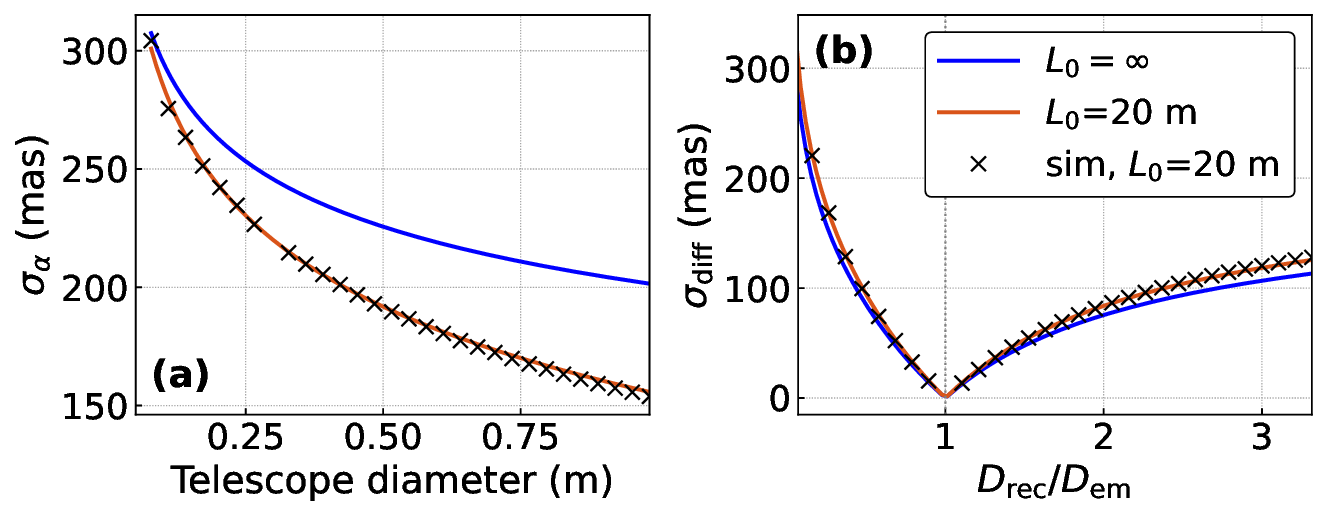}
\caption{(a) TT angular jitter $\sigma_\alpha$ generated by atmospheric turbulence vs. telescope diameter, for $r_0=15$\,cm and $L_0=20$\,m, comparing Kolmogorov and von K\'arm\'an models. (b) Differential TT signal amplitude $\sigma_\mathrm{diff}$ (Eq.~\ref{eq:diff_tt}) vs. $D_\mathrm{rec}/D_\mathrm{em}$; the null at $D_\mathrm{rec} = D_\mathrm{em} = 0.3$\,m reflects reciprocity. Solid lines: analytical model; symbols: OOPAO simulations.}
\label{fig:diff}
\end{figure}

The TT signals measured with different aperture diameters are strongly but not perfectly correlated (Eq.~\ref{eq:Wilson}): the correlation decreases as the ratio $D_\mathrm{rec}/D_\mathrm{em}$ departs from unity, and a finite outer scale $L_0$ further reduces it with respect to the Kolmogorov case. This partial decorrelation is the physical resource exploited by our method, as it provides complementary information about the turbulent wavefront. The resulting TT angular jitter measured by the receiver, containing both the uplink and downlink contributions, follows from Eq.~\ref{eq:diff_tt}.
\begin{equation}
\sigma_{\mathrm{diff}}
=
\sqrt{
\sigma_{1}^{2}
+
\sigma_{2}^{2}
-
2\,\sigma_{1}\sigma_{2}\,
\rho_{\mathrm{TT}}(D_{1},D_{2})
},
\label{eq:diff_tt}
\end{equation}
In Fig.~\ref{fig:diff}(b), the differential signal vanishes when $D_{\mathrm{rec}} = D_{\mathrm{em}}$ (reciprocity), and grows as the receiver and emitter apertures diverge, more steeply for the von K\'arm\'an model due to its faster decorrelation. The agreement between the analytical model and the OOPAO simulations over the full tested range --- which also holds for the individual variances (Fig.~\ref{fig:diff}(a)) --- validates the theoretical description, including, implicitly, the correlation $\rho_\mathrm{TT}$ of Eq.~\ref{eq:Wilson} from which $\sigma_\mathrm{diff}$ is built. We note that aperture size diversity provides no information precisely at reciprocity, where $\sigma_\mathrm{diff}=0$; this limit could be complemented by combining our approach with the time-delay technique of \cite{ragazzoni1996propagation}, which could naturally be incorporated in the LMMSE framework through the lagged input vector introduced in Sect.~\ref{sec:LMMSE}.

\subsubsection{Differential uplink-downlink signal and uplink signal}

\begin{figure}[htb]
\centering
\includegraphics[width=0.6\linewidth]{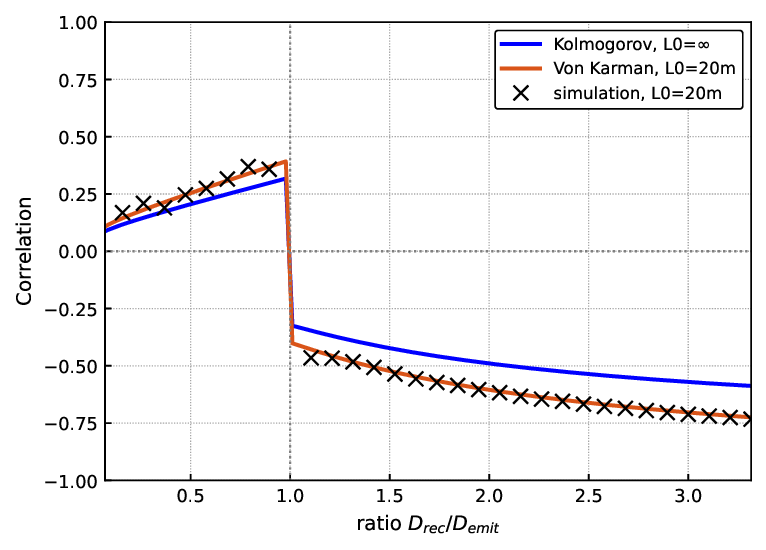}
\caption{Correlation $\rho_{R,\mathrm{up}}$ between the TT measured by the receiver (uplink+downlink) and the uplink beam, as a function of receiver diameter ($D_\mathrm{em} = 0.3$\,m).}
\label{fig:corr_uplink}
\end{figure}

The previous sections considered only instantaneous spatial statistics between concentric apertures. We now relate the instantaneous differential measurement to the true uplink TT signal itself, which motivates the spatio-temporal estimation framework introduced in Sect.~\ref{sec:LMMSE}. We propagate a cylindrical beam with diameter equal to the emitter aperture (30\,cm) through a turbulent phase screen and estimate the resulting TT in mas. We then propagate a second beam whose diameter matches the chosen receptor size through the same turbulence realization and compute the differential TT. In Eq.~\ref{eq:corr_residual_uplink}, we give the associated formula.
The receiver measures the differential signal $R(t) = s_{\mathrm{down}}(t) - s_{\mathrm{up}}(t)$, where $s_{\mathrm{up}}(t)$ and $s_{\mathrm{down}}(t)$ are the instantaneous uplink and downlink tip-tilt time series. Its correlation with the true uplink signal is given by
\begin{equation}
\rho_{R,\mathrm{up}}(D_{\mathrm{em}},D_{\mathrm{rec}})
=
\frac{
\rho_{\mathrm{TT}}(D_{\mathrm{em}},D_{\mathrm{rec}})\,\sigma_{\mathrm{down}} - \sigma_{\mathrm{up}}
}{
\sigma_{\mathrm{diff}}
},
\label{eq:corr_residual_uplink}
\end{equation}
where $\sigma_{\mathrm{up}}$ and $\sigma_{\mathrm{down}}$ are the standard deviations of $s_{\mathrm{up}}(t)$ and $s_{\mathrm{down}}(t)$, the instantaneous uplink and downlink tip-tilt time series for the emitter (diameter $D_{\mathrm{em}}$) and receiver (diameter $D_{\mathrm{rec}}$) apertures respectively, $\rho_{\mathrm{TT}}(D_{\mathrm{em}},D_{\mathrm{rec}})$ is given by Eq.~\ref{eq:Wilson}, and $\sigma_{\mathrm{diff}}$ is the differential signal amplitude defined in Eq.~\ref{eq:diff_tt}. At reciprocity ($D_{\mathrm{em}}=D_{\mathrm{rec}}$), $\rho_{R,\mathrm{up}}$ is undefined since $\sigma_{\mathrm{diff}}=0$; in the limit of a large receiver ($D_{\mathrm{rec}} \to \infty$), $\rho_{R,\mathrm{up}} \to -1$.
Figure~\ref{fig:corr_uplink} shows the correlation computed for an emitter with a fixed diameter of 30\,cm. The correlation coefficient changes sign when the emitter diameter equals the receptor diameter. When the emitter aperture is larger than the receptor aperture, the TT is dominated by the downlink signal. Conversely, when the emitter aperture is smaller, the downlink contribution becomes negligible and the differential TT is dominated by the uplink term. Also, the impact of a finite outer scale suppresses the large scale turbulent structures that TT is most sensitive to, and which are responsible for the shared component between mismatched apertures; their removal accelerates the decorrelation between emitter and receiver, making the residual signal more anti-correlated with the true uplink. This behavior underlies the spatial contribution to the estimation discussed in Sect.~\ref{sec:results}: apertures increasingly mismatched from the emitter carry a differential signal increasingly correlated with the uplink, which is consistent with the decrease of the estimation error with the maximum receiver diameter (Fig.~\ref{fig:map}). Finally, the central idea is to establish a functional relationship between the uplink TT and the set of differential uplink-downlink measurements obtained for multiple receptor diameters.

\subsection{Linear predictor}\label{sec:LMMSE}

The estimation of the uplink TT is obtained by combining multi-aperture measurements through a linear predictor to estimate the uplink. Let $\mathbf{x}(t) \in \mathbb{R}^N$ denote the vector of differential uplink-downlink TT measurements from the set of concentric apertures, and $y(t)$ the true uplink TT. To account for temporal dynamics, we define a lagged input vector:
\begin{equation}
\mathbf{x}_{\mathrm{lag}}(t) =
\bigl[x_1(t{-}1),\ldots,x_N(t{-}1),\,x_1(t{-}2),\ldots,x_N(t{-}n_\mathrm{lag})\bigr]^\top,
\label{eq:extended}
\end{equation}
showing explicitly how temporal information and multiple aperture measurements are concatenated into a single feature vector. The estimator thus draws on two distinct information channels: (i) the \emph{spatial} channel, provided at each instant by the $N$ partially decorrelated aperture measurements, which is the enabling resource of the method; and (ii) the \emph{temporal} channel, provided by the last $n_\mathrm{lag}$ frames, which exploits the temporal correlation of the turbulent tilt to refine the estimate. Their respective contributions are quantified in Sect.~\ref{sec:results}. The LMMSE predictor \cite{bishop2006pattern} is:
\begin{equation}
\hat{y}(t) = \mathbf{w}^\top \mathbf{x}_{\mathrm{lag}}(t),
\qquad
\mathbf{w} = \mathbf{C}_{xx}^{-1} \mathbf{C}_{xy},
\label{eq:linear}
\end{equation}
where $\mathbf{C}_{xx} = \mathbb{E}[\mathbf{x}_\mathrm{lag}\mathbf{x}_\mathrm{lag}^\top]$ is the covariance matrix of the measurements and $\mathbf{C}_{xy} = \mathbb{E}[\mathbf{x}_\mathrm{lag}\,y]$ the cross-covariance vector. In practice, these quantities are estimated from telemetry data, leading to the empirical regularized solution:
\begin{equation}
\mathbf{w} = \left(\mathbf{X}^\top \mathbf{X} + \lambda \mathbf{I} \right)^{-1} \mathbf{X}^\top \mathbf{y},
\label{eq:tiko}
\end{equation}
where $\mathbf{X}$ is the lagged data matrix and $\lambda$ a regularization parameter accounting for noise and finite sample effects. The LMMSE predictor exploits both the spatial diversity provided by multiple apertures and the temporal correlations of the signal to optimally reconstruct the uplink TT.

\section{Results}\label{sec:results}

The LMMSE estimator is trained over 2000 iterations with a time step of 2\,ms using the parameters in Table~\ref{tab:params}, then evaluated on 8000 independent steps. The single-layer case establishes baseline performance under the simplified framework; the two-layer case includes a high-altitude layer at 8\,km and the associated LGS cone effect. In the two-layer configuration, the $C_n^2$ fractional weights are 75\% for the ground layer and 25\% for the 8\,km layer, for the same integrated $r_0=15$\,cm. This altitude and weighting are representative of the strong free-atmosphere layer associated with the jet stream in standard site profiles \cite{schock2009tmt}.

\begin{table}[htb]
\centering
\caption{Simulation parameters}
\begin{tabular}{lcc}
\hline
Parameter & 1 layer & 2 layers \\
\hline
$r_0$ / $L_0$ & \multicolumn{2}{c}{15\,cm / 20\,m} \\
Altitudes (km) & 0 & 0, 8 \\
$C_n^2$ fraction (\%) & 100 & 75, 25 \\
Wind speed (m\,s$^{-1}$) & 10 & 10, 15 \\
Emitter diam. (m) & \multicolumn{2}{c}{0.3} \\
Receiver diam. (m) & \multicolumn{2}{c}{0.26--0.98 (12 values)} \\
Time step (ms) & \multicolumn{2}{c}{2} \\
Temporal lags $n_\mathrm{lag}$ & \multicolumn{2}{c}{30 (60\,ms)} \\
\hline
\end{tabular}
\label{tab:params}
\end{table}

\begin{figure}[htb]
\centering
\includegraphics[width=0.6\linewidth]{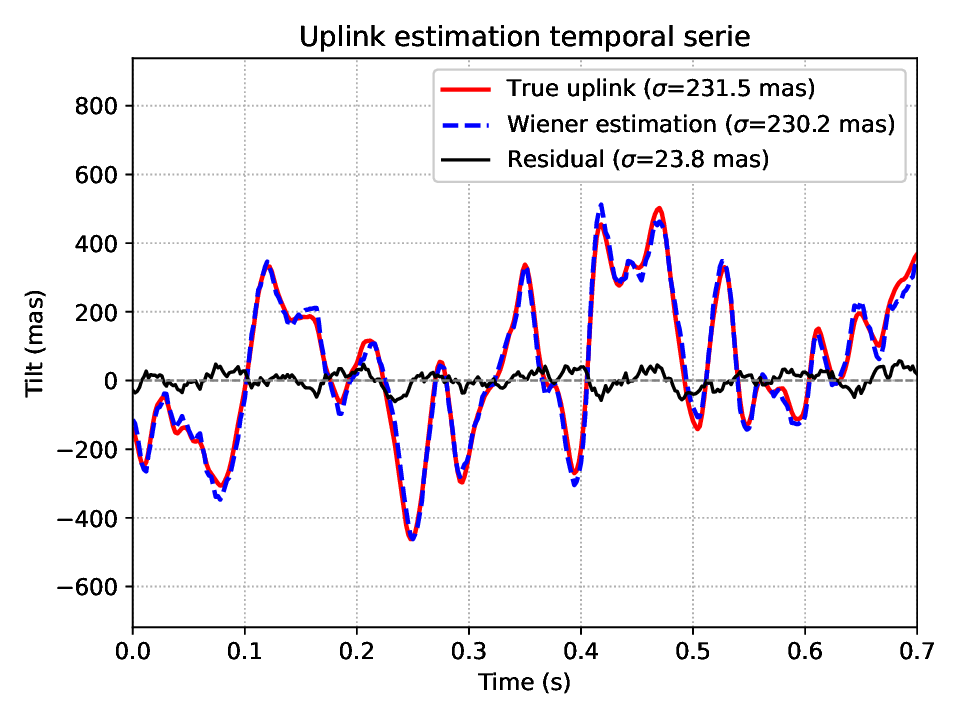}
\caption{Temporal sequence of estimated tilt vs. true tilt, single-layer configuration. Each time step represents 2\,ms.}
\label{fig:temporal}
\end{figure}

After training, we only use data from the LGS to estimate the uplink TT signal. We compare the estimated values to the true values in Fig.~\ref{fig:temporal} over a subset of 350 steps in the single-layer configuration. The standard deviation of the open-loop uplink is $\sigma_\mathrm{OL}=232$\,mas and the standard deviation of the error signal is $\sigma_\mathrm{err}=24$\,mas, representing a 90\% reduction. To verify that the reconstructor generalizes beyond the specific turbulence realization used for training, we applied the filter trained on a single seed to ten additional, independent turbulence realizations (same atmospheric parameters), obtaining an average error of $24\pm2$\,mas, confirming that the reconstructor captures the underlying physical covariance structure rather than overfitting to a particular turbulence sequence. Adding the second turbulent layer, including the associated cone effect on the downlink beam, degrades the estimation to $\sigma_\mathrm{err}=34$\,mas.

\begin{figure}[htb]
\centering
\includegraphics[width=0.7\linewidth]{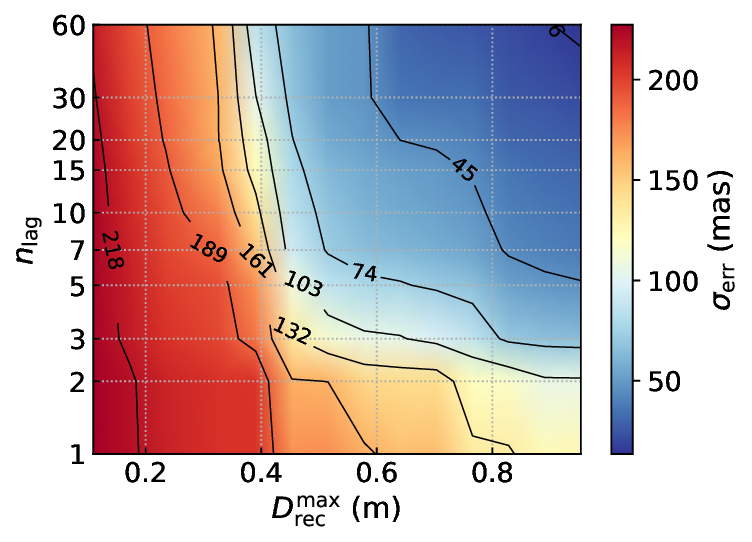}
\caption{Estimation error $\sigma_\mathrm{err}$ as a function of the maximum receiver diameter $D_\mathrm{rec}^\mathrm{max}$ and the temporal history length $n_\mathrm{lag}$, single-layer configuration. Contours are labelled in mas.}
\label{fig:map}
\end{figure}

The two information channels exploited by the estimator, and the two corresponding implementation constraints, are summarized in Fig.~\ref{fig:map}, which maps $\sigma_\mathrm{err}$ over the plane spanned by the maximum receiver diameter $D_\mathrm{rec}^\mathrm{max}$ and the temporal history length $n_\mathrm{lag}$. Both channels contribute substantially: at the largest receiver diameter, increasing $n_\mathrm{lag}$ from 1 to 60 reduces the error by a factor of about 5, from 114\,mas to 24\,mas, with the improvement slowing markedly beyond $n_\mathrm{lag}\simeq15$ (30\,ms). Along the $D_\mathrm{rec}^\mathrm{max}$ axis, the error decreases monotonically with receiver size, from above 213\,mas at 0.3\,m to below 24\,mas at $\simeq1$\,m, as anticipated from the growth of $|\rho_{R,\mathrm{up}}|$ in Fig.~\ref{fig:corr_uplink}. Aperture diversity is the enabling resource of the method --- no differential signal exists without it --- while the temporal channel accounts for most of the remaining gain. The performance reported here requires $D_\mathrm{rec}^\mathrm{max}\simeq1$\,m for a 0.3\,m emitter.

A dedicated tip-tilt mirror inserted in the LGS optical path could provide an effective means of correcting this component in real time using only LLT information. In a multi-LGS system, each LGS is launched from its own LLT, mounted around the aperture of the science telescope, and each LLT independently estimates and corrects its own uplink TT using the proposed method. Because each uplink beam propagates through a distinct turbulent volume, the residual uplink contribution left after this correction is statistically uncorrelated from one LGS to another --- unlike the downlink TT, which is common to all LGS through the shared science-path dominated by ground-layer turbulence. When the $N$ LGS wavefront sensor measurements are combined in the low-order reconstruction of the science path, this common downlink signal is essentially unchanged, while the uncorrelated residual uplink contributions average down as $1/\sqrt{N}$. This averaging is only effective if the downlink paths of the $N$ LGS retain sufficient overlap through the turbulent volume; the asterism geometry should therefore be chosen to maximize this overlap at the altitude of the dominant turbulent layer. For a classical four-LGS system, the final one-axis TT angle is $\sigma \approx 34/\sqrt{4} = 17$\,mas.

To put this into context, we use the expression from \cite{hardy1998adaptive} for the total PSF core diameter including the diffraction-limited PSF and additional jitter $\sigma_\alpha$:
\begin{equation}
a_c = 1.22 \frac{\lambda}{D}
\left[ 1 + 5.17 \left( \frac{D}{\lambda} \right)^{2} \sigma_{\alpha}^{2} \right]^{1/2}.
\label{eq:hardyFWHM}
\end{equation}
In Table~\ref{tab:fwhm}, we compare the diffraction-limited PSF core diameter and the residual TT contribution in $H$-band for four representative telescope diameters, considering $\sigma_\mathrm{err}=17$\,mas. The corrected core diameter stays within a factor 1.02--4.6 of the diffraction limit across the 2--40\,m range, and remains a substantial improvement over the seeing-limited FWHM ($2.3''$), demonstrating that this method provides a significant gain in fields where no NGS is available.

\begin{table}[htb]
\centering
\caption{$H$-band PSF core diameter for a four-LGS system ($r_0=15$\,cm)}
\begin{tabular}{lcccc}
\hline
Telescope diam. (m) & 2 & 4 & 8 & 40 \\
\hline
Diffraction limit (mas) & 208 & 104 & 52 & 10 \\
+ 17\,mas residual TT (mas) & 213 & 114 & 70 & 48 \\
Ratio to diffraction limit & 1.02 & 1.10 & 1.35 & 4.6 \\
\hline
\end{tabular}
\label{tab:fwhm}
\end{table}

\section{Conclusion}

We have demonstrated that up to 90\% of the uplink TT can be retrieved
by exploiting the combined uplink and downlink signals in a monostatic
laser guide star configuration, using telemetry from the laser launch
telescope alone. Combined with the averaging over multiple
LGSs, the residual jitter (17\,mas for a four-LGS system) remains well
below the diffraction limit of 1--10\,m class telescopes in the near
infrared, for which the method is therefore particularly promising.
This approach could ultimately relax the need for natural guide stars
and enable adaptive optics correction across the entire sky, and is
equally relevant to AO pre-compensation of ground-to-space optical
communication uplinks \cite{lognone2024tip,barrett2026tip}. The
complementarity with the time-delay technique noted in Sect.~\ref{sec:method}
could further extend its applicability near reciprocity. Future work must address realistic
implementation challenges, including (1) the development of a dedicated
wavefront sensor providing simultaneous TT measurements across multiple
aperture sizes, (2) the propagation of measurement noise through the LMMSE
estimator, and (3) validation under varying and multi-layer atmospheric
conditions. These preliminary results establish the fundamental
viability of aperture size diversity as a means to mitigate the
long-standing LGS TT indetermination problem.

\subsection*{Acknowledgments}
Simulations were performed using the OOPAO open-source adaptive optics simulation framework.

\subsection*{Disclosures}
The authors declare no conflicts of interest.

\subsection*{Funding}
C.C. acknowledges support from FCT with ref 2024.16728.PEX.

\subsection*{Data availability}
Data underlying this paper are available from the authors upon reasonable request.

\bibliographystyle{unsrtnat}
\bibliography{references}

\end{document}